\documentclass[aps,prd,preprint,nofootinbib]{revtex4}
\usepackage{epsf,epsfig,graphics,graphicx}
\usepackage{verbatim,color,ulem}
\newcommand{\be}{\begin{equation}}
\newcommand{\ee}{\end{equation}}
\newcommand{\bea}{\begin{eqnarray}}
\newcommand{\eea}{\end{eqnarray}}
\newcommand{\ba}{\begin{array}}
\newcommand{\ea}{\end{array}}
\usepackage{amssymb,amsmath,amsthm}
\usepackage[mathscr]{eucal}
\usepackage{enumerate,color,verbatim,multirow,comment}
\usepackage[caption=false]{subfig}

\begin{document}

\title{Shock Waves in Holographic EPR pair}

\author{Chen-Pin Yeh}
\email{chenpinyeh@gms.ndhu.edu.tw}
\affiliation{Department of Physics, National Dong Hwa University, Hualien, Taiwan, R.O.C.
}

\begin{abstract}
We study real-time correlators for $N=4$ super Yang Mill fields coupled to a pair of entangled quarks using holography, in the setup that energy quanta sent from one quark perturb the quantum state of the fields and affect the other quark. We make the connection with the ER=EPR conjecture by considering the situation when two quarks are uniformly accelerating opposite to each other. The dynamics of quarks, in the gravity dual, is described by the string worldsheet theory, which in this case has the induced metric describing a two-sided AdS black hole, or a wormhole. Energy quanta sent by one of the quarks produce the shock wave on the worldsheet. We find the effect of shock wave on the boundary field correlators and we discuss the consequence for the ER=EPR conjecture.

\end{abstract}

\maketitle

\section{introduction}
The AdS/CFT correspondence is an example of holographic duality saying that the gravity theory in anti de Sitter (AdS) background is equivalent to $N=4$ super Yang Mill theory (a CFT) on the AdS boundary. The correspondence geometrizes the quantities in quantum mechanics, just like general relativity geometrizes the quantities in classical physics. While general relativity provides insights into the gravity force and answers questions that are difficult to explain in Newtonian gravity, we expect the AdS/CFT to guide us to the theory of quantum gravity, and resolve paradoxes in current understanding of quantum gravity, in particular the information problem in black hole formation and evaporation. Previous studies of the information problem have shown a close relation between areas in gravity and information in quantum theory, and it formed the basis of the holographic principle \cite{tHooft_93,Susskind_94}. The AdS/CFT, in particular, identifies the boundary CFT entanglement entropy with the area of a certain minimal surface in AdS gravity\cite{Ryu_06}.

ER=EPR is another interesting geomtrization of quantum entanglement, which draws on an analogy between Einstein-Podolsky-Rosen pairs (EPR pairs) and Einstein-Rosen bridges (ER bridges or wormholes)\cite{ER=EPR}. In the gravity theory, wormholes are classical solutions of general relativity, that  some nonlocal connection between two space-time regions is created, yet allow no superluminal signals transmission (hinge on some energy conditions). Similarly, in EPR pairs, non-local quantum entanglement correlates two particles. However careful study of quantum measurement also revealed no superluminal signal transmission. Moreover, entanglement entropy of EPR pair is related to the throat areas of wormholes. Thus we expect that when a wormhole's throat pinches, the EPR pair starts to disentangle. A naive question to ask in this process is how the entanglement encodes the dynamics of quantum wormhole. Or we can also ask how different interaction between entanglement pairs changes the geometry. The current work is motivated by such questions.  To make the setting more precise, we consider a probe string in AdS space with two ends anchored on the boundary, which in the content of AdS/CFT, is dual to the quark and anti-quark EPR pair moving through $N=4$ super Yang-Mills fields. As two ends of the string are uniformly accelerating opposite to each other along the boundary, the induced metric on the string worldsheet is exactly the one for a two-sided AdS black hole (a wormhole) as will be seen later in this work. It was first noticed in \cite{Jensen_13} and \cite{Sonner_13} that in this case, there is also a close relation between worldsheet wormhole and the entanglement of the quarks in the spirit of ER=EPR. For example, the entanglement entropy of the EPR pair is equal to the black hole entropy, and the quark anti-quark pair creation rate is equal to the onshell string worldsheet action. In \cite{Pedraza_13}, the relation between the EPR pairs and the existence of wormholes has been extended to more general quark trajectories in this probe string setup. Furthermore, they also argued that the behind-horizon regions in worldsheet wormholes encodes the information of gluon radiations from quark pairs. But we still don't know how the change of trajectories of the EPR pairs can be detected in the wormhole geometries (and vise versa). The simplest setup to examine this question is by considering shock waves on the worldsheet wormholes, which correspond to the energy transfer between entangled quarks in the boundary theory. These shock waves, in the case when the end points of string are uniformly accelerating were considered in \cite{Murata_17}\footnote{The setup and method used in this paper is the same as the ones in \cite{Shenker_13}, where they consider shock waves in two-sided BTZ black holes.}, and the field correlators between two wormhole boundaries are derived using the geodesic approximation in AdS/CFT. They found that the fields coupled to left and right quark have equal time correlators decaying with time, and the effect of shock wave is not important at late times.

Certainly it will be interesting to probe the change of wormhole geometries directly by the entanglement entropy\footnote{It may not be able to completely specify the wormhole geometry by entanglement entropy only. For example, the complexity is also needed\cite{Lenny_14}.}. One difficulty is that when the classical wormhole description is valid, the CFT is strongly-coupled and it is ambiguous to separate quarks and gluons degrees of freedom (see \cite{Jensen_13,Hatta_11,Garcia_13,Agon_14,Grein_14} for detailed discussion). So unlike the case in \cite{Shenker_13}, where they can examine the effect of shock waves by calculating the mutual information between fields in two boundaries regions, using the Ryu-Taganayagi formula \cite{Ryu_06}, here the mutual information between two quarks is difficult to define. In the current work, we will use the holographic influence functional method developed in \cite{Lee_15} (and references therein) to evaluate the field correlators that couple to quarks in this shock wave background. The advantage of our method is that we directly obtain the real-time field correlators. The formalism also makes it easier to study the quark dynamics, by deriving the Langevin equation for the quark positions. In \cite{Lee_19}, the quark correlators derived in this way have been used to calculate the time evolution of entanglement entropy between the quark and fields, and it was found that, at late times, entanglement entropy is mainly determined by the field correlators. Thus we expect the field correlators obtained this way give qualitative behaviors of the quark dynamics and its entanglement entropy. But for detailed time-dependence, solving the coupled Langevin equations is still required.

In \cite{Kawamoto_22}, the method of the holographic influence functional has been extended to the case of two-end string, which is what relevant for current study. Using this method we find that the effect of the shock wave on field correlators is perturbatively determined by the small parameter $\gamma$ in the double scaling limit to be defined later. The shift of the wormhole horizon after the shock wave passes is proportional to $\gamma$. In particular when $\gamma<0$, the shock wave renders the wormhole traversable. We find that at late times $t\gg \frac1{T}$, where $T$ is the temperature of the black hole, the effect of the shock wave on correlators becomes negligible. When $t\ll \frac1{T}$ but larger than the UV cutoff scale\footnote{The quark mass is also determined by the scale $z_m^{-1}$, which is assumed to be large.} $z_m$, we find that the cross correlators between fields coupled to the entangled quark pair behave differently depending on whether $\gamma$ is positive or negative. When $\gamma>0$, two ends of the string remain causally disconnected, and the cross correlators decreases more rapidly comparing to the case with no shock wave. As $\gamma<0$, two quarks become causally connected from the worldsheet point of view (even though in boundary two quarks are still out of causal contact), and the cross correlators increase for a while (of the scale of wormhole throat size, $\frac{\gamma}{T}$) and then decrease eventually\footnote{In the free field theory, the same problem without the shocks can be solved directly using the Gaussian reduced density matrix\cite{Hu_10}. In some parameter regions, they also found the buildups of cross correlators and entanglement entropy between two objects accelerating opposite to each other.}. The implications of these behaviors on ER=EPR will be discussed in the final section. In section II, we will review the induced worldsheet backgrounds for probed accelerating string moving in AdS space. In section III, we patch the worldsheets with different acceleration together to form a shock wave background. Then in section IV, we use the holographic influence functional in this shock wave background to derive the backreacted real-time field correlators and analyze their early-time and late-time behaviors. We also discuss the relation of these field correlators to the out-of-time-order correlators(OTOCs). In section V, we consider the case with two shock waves, and then we conclude and point out some future directions in section VI.

\section{accelerating string}
\begin{figure}[h]
\centering
\includegraphics[scale=0.7]{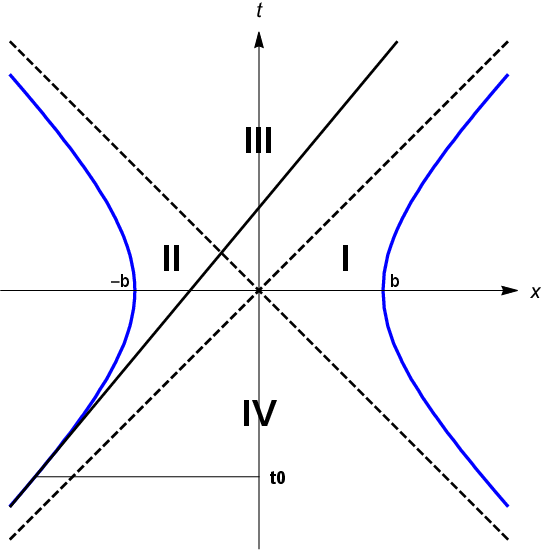}
\caption{A light ray emitted from the left quark at $t_0$ (solid black lines) in the worldsheet. The hyperbolic curves (blue) are trajectories of quark and anti-quark(we set $z_m=0$ here). Two quarks are causally disconnected with the horizons at $t=\pm x$ (dash lines). Two horizons divide the worldsheet into four regions, where we label I,II,III and IV.}
\label{lightray}
\end{figure}

In this section, we review the worldsheet theory of a string moving in $AdS_{d+1}$ with its two ends uniformly accelerating opposite with each other in the boundary of $AdS_{d+1}$. It turns out that the dimension $d$ will not play any role in the following discussion. We consider the $AdS_{d+1}$ metric in the Poincar\'{e} coordinates given by
  \be
  \label{ads}
  ds^2=\frac{R^2}{z^2}(-dt^2+dz^2+dx^2+\sum_{i=1}^{d-2}dy_i^2)\, ,
  \ee
where $R$ is the curvature radius. The probed string in this background is described by the Nambu-Goto action
\begin{equation}
\label{action}
S=-T_0 \int d\tau d\sigma \sqrt{- h} \, ,
\end{equation}
where $T_0$ is the string tension and $h=\text{det} \, h_{ab}$ is the determinant of the induced metric. We choose a worldsheet parameterization, $(\tau,\sigma)=(t,z)$ and the embedding of the string as $X^{\mu}(t,z)=(t,z,x(t,z),0,\cdots,0)$ with $\mu=0, 1, 2,\cdots, d$.
Then, $\sqrt{-h}=\frac{R^2}{z^2} \sqrt{1-\dot x^2+x'^2}$.
The classical equation of motion is given by
\be
\label{eom}
\frac{\partial}{\partial t} \Big(\frac{\dot x}{z^2 \sqrt{1-\dot x^2+x'^2}}\Big)-\frac{\partial}{\partial z} \Big(\frac{ x'}{z^2 \sqrt{1-\dot x^2+x'^2}}\Big)=0 \, ,
\ee
where $'=\frac{\partial}{\partial z}$ and $\cdot=\frac{\partial}{\partial t}$.
This equation has exact solutions \cite{Xiao_08},
  \be
  \label{xb}
  x_b(t,z)=\pm\sqrt{t^2+b^2-z^2} \, ,
  \ee
parameterized by a real constant $b$. Note that later we will impose a boundary cutoff at $z=z_m$ with $z_m\ll b$, and identify two string end points, $x_b(t,z_m)=\pm\sqrt{t^2+b^2-z_m^2}$ as the trajectories of the entangled quark pair. These are exactly the trajectories of a point particle moving with constant acceleration $\frac1{\sqrt{b^2-z_m^2}}$ in $\pm x$ directions\footnote{This interpretation can also be confirmed by studying the Unruh effect on these quarks\cite{Caceres_10}.}. The induced metric on the string worldsheet is \cite{Kawamoto_22}
    \be
    \label{ws}
    \frac{1}{(t^2+b^2-x^2)^2}\left((x^2-b^2)dt^2-2txdtdx+(t^2+b^2)dx^2\right)\, .
    \ee
We call this $(x,t)$-coordinate the laboratory frame. The remaining discussion in this section is to relate this laboratory frame to the Kruskal-like frame, in which the calculation of field correlators is more convenient. But the physical interpretation is in the laboratory frame. To do the coordinate transform to light-like coordinates, note that the light rays on the wordsheet follow the tangent lines to the trajectories of the end points of string. This can be seen directly from the null geodesics in (\ref{ws}), and those null lines coincide at the string end points with the boundary ($AdS_{d+1}$) null geodesics. So we can parameterize the light rays by $t_0$, which is the laboratory time when they are emitted from or reach the boundaries \cite{Jensen_13},
  \be
  \label{rays}
  x(t)=\pm\frac{tt_0+b^2}{\sqrt{t_0^2+b^2}}\, .
  \ee
 $x(t)$'s here are the tangent lines to the quark trajectories on $(x,t)$-plane. The casual structure is drawn in figure \ref{lightray}, where we see that two quarks are causally disconnected and there are horizons at $t=\pm x$. We divide the worldsheet into four regions (I, II, III, and IV) separated by the horizons.
The metric (\ref{ws}) describes a two-sided $AdS_2$ black hole as suggested by the casual structure, and to see this we use the light-like coordinates, $u$ and $v$ (from the inverse relations in (\ref{rays})). In region I, we set
 \be
 u=\frac{b^2t+xb\sqrt{b^2+t^2-x^2}}{x^2-t^2},~~~v=\frac{b^2t-xb\sqrt{b^2+t^2-x^2}}{x^2-t^2} \, ,
 \ee
with $-\infty<v<u<\infty$. Then the worldsheet metric becomes
 \be
 ds^2=\frac{2(-b^2+uv+\sqrt{b^2+u^2}\sqrt{b^2+v^2})}{\sqrt{b^2+u^2}\sqrt{b^2+v^2}(u+v)^2}dudv \, .
 \ee
To put the metric in a more standard $AdS$ black hole form, we further set $\tilde{u}=b\sinh^{-1}\frac{u}{b}$ and $\tilde{v}=b\sinh^{-1}\frac{v}{b}$, then we have
  \be
  ds^2=-\frac{d\tilde{u}d\tilde{v}}{b^2\sinh^2\frac{\tilde{u}-\tilde{v}}{2b}} \, ,
  \ee
which is exactly the metric for the outside region of $AdS_2$ black hole with temperature $T=\frac{1}{2\pi b}$. We can do the similar coordinate transform in region III, by setting\footnote{Note that, here and later, to avoid too many indices, we use the same notation for coordinates in different regions.}
  \be
  \label{uv}
 u=\frac{-b^2t-xb\sqrt{b^2+t^2-x^2}}{x^2-t^2},~~~v=\frac{b^2t-xb\sqrt{b^2+t^2-x^2}}{x^2-t^2} \, ,
 \ee
 with $-\infty<v<u<\infty$. Then in $\tilde{u}$ and $\tilde{v}$ coordinates, the metric in region III becomes
   \be
   ds^2=\frac{d\tilde{u}d\tilde{v}}{b^2\cosh^2\frac{\tilde{u}-\tilde{v}}{2b}} \, ,
   \ee
 which is the metric for $AdS_2$ black hole interior. The coordinate transforms in region II (IV) are related to region I (III) by setting $x\rightarrow -x$ and $t\rightarrow -t$. We see that the four regions cover the whole worldsheet (\ref{ws}) and is exactly the $AdS$ eternal black hole, and its AdS/CFT interpretation is given in \cite{Maldacena_01}. The relation between $(\tilde{u},\tilde{v})$-coordinate and $(x,t)$-coordinate is useful later. In region I, we have
  \be
  \label{uvti}
  t=\frac{b\left(\cosh\frac{\tilde{u}}{b}-\cosh\frac{\tilde{v}}{b}\right)}{\sinh\frac{\tilde{u}-\tilde{v}}{b}},~~x=\frac{b\left(\sinh\frac{\tilde{u}}{b}-\sinh\frac{\tilde{v}}{b}\right)}{\sinh\frac{\tilde{u}-\tilde{v}}{b}} \, .
  \ee
Thus on the worldsheet, we have
  \be
   \label{z}
  z=\sqrt{b^2+t^2-x^2}=b\tanh\frac{\tilde{u}-\tilde{v}}{2b},~~~\frac{t}{x}=\tanh\frac{\tilde{u}+\tilde{v}}{2b} \, .
  \ee
The same relations hold also in region II. In region III and IV, we have the relations
  \be
  z=\sqrt{b^2+t^2-x^2}=b\coth\frac{\tilde{u}-\tilde{v}}{2b},~~~\frac{t}{x}=\coth\frac{\tilde{u}+\tilde{v}}{2b} \, .
  \ee

\begin{figure}[h]
\centering
\includegraphics[scale=1]{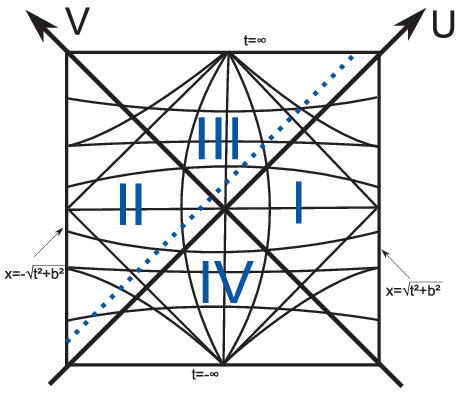}
\caption{The worldsheet metric in $(U,V)$-coordinate. Both $U$ and $V$ have their ranges from $-b\pi$ to $b\pi$. The left (right) boundary is identified with the left (right) quark trajectory, and is described by $U-V=-b\pi$ ($U-V=b\pi$). We show the constant $t$ and $x$ lines (only schematically) and the four regions I, II, III, IV of the worldsheet. Here the (blue) dash line is a constant $\tilde{u}$ line.}
\label{conformal}
\end{figure}

To see the four regions patched in $AdS_2$ global coordinates, we can make the following coordinate transforms. First, we can define $u'=2b\tanh\frac{\tilde{u}}{2b}$ and $v'=2b\tanh\frac{\tilde{v}}{2b}$ (thus we have $-2b<v'<u'<2b$). And in region I and II, we have the metric,
   \be
   \frac{-4du'dv'}{(u'-v')^2} \, .
   \ee
Here the boundaries are located at $u'=v'$ and the black hole horizons are at $u'=2b$ and $v'=-2b$. In region III and IV, we have the metric
   \be
     \frac{16b^2du'dv'}{(-4b^2+u'v')^2} \, .
   \ee
And $u'=v'$ corresponds to $t\rightarrow\infty$ ($t\rightarrow -\infty$) in region III (region IV). These coordinates cover the Poincar\'{e} patch of $AdS_2$. We can embed the Poincar\'{e} patch in the global $AdS_2$ by using the coordinates, $u''=2b\tan^{-1}\frac{u'}{2b}$ and $v''=2b\tan^{-1}\frac{v'}{2b}$, and then we have metric in region I and II as
  \be
   \frac{-du''dv''}{b^2\sin^2\frac{u''-v''}{2b}} \, ,
  \ee
where now the black hole horizons are at $u''=\frac{b\pi}{2}$ and $v''=-\frac{b\pi}{2}$, and the Poincar\'{e} horizons are at $u''=b\pi$ and $v''=-b\pi$. And in region III and IV, we have the metric as
   \be
   \label{uvprime}
   \frac{du''dv''}{b^2\cos^2\frac{u''+v''}{2b}} \, .
  \ee
Finally, we can define the Kruskal-like coordinates $U$ and $V$ that covers all four regions of the worldsheet.  In region I, $U=v''+\frac{b\pi}{2}$, $V=u''-\frac{b\pi}{2}$ (thus $b\pi>U>0, -b\pi<V<0$), in region II, $U=-v''-\frac{b\pi}{2}$, $V=-u''+\frac{b\pi}{2}$ (thus $-b\pi<U<0, b\pi>V>0$), in region III, $U=v''+\frac{b\pi}{2}$, $V=-u''+\frac{b\pi}{2}$ (thus $b\pi>U>0, b\pi>V>0$), and in region IV, $U=-v''-\frac{b\pi}{2}$, $V=u''-\frac{b\pi}{2}$ (thus $-b\pi<U<0,-b\pi<V<0$), then two boundaries are at $U-V=\mp b\pi$ and the horizons at $U=0$ and $V=0$. Note that in all four regions, we have the metric,
  \be
  \label{UV}
   \frac{-dUdV}{b^2\cos^2\frac{U-V}{2b}} \, .
  \ee
The worldsheet in $(U,V)$-coordinate is drawn in figure \ref{conformal}.

\section{the shock wave background}
Here we consider the shock wave configuration and its matching condition in region II. It can be easily generalized to other regions by the coordinate relations described in previous section. These shock wave solutions were found in \cite{Murata_17} (see\cite{Mikhailov_03,Hubeny_14,Vegh_15}, for the discussion of more general solutions). For convenience, here we describe the solutions using $(\tilde{u},\tilde{v})$-coordinates. Certainly, the geometry will not depend on the coordinates. However later we will take the so-called double scaling limit of the shock wave geometry, and it should be noticed that the double scaling limit does not commute with the coordinate transformation.
The shock wave we consider is emitted from the left quark at the time $t=t_0$ follows the light-like trajectory(\ref{rays}), and in $(\tilde{u},\tilde{v})$-coordinates, the shock wave then follows the constant $\tilde{u}$ trajectory, $\tilde{u}=\tilde{u}_0\equiv -b\sinh^{-1}\frac{t_0}{b}$. Note that $-\tilde{u}_0$ is exactly the proper time along the quark trajectory corresponds to the laboratory time $t_0$ and the shock wave trajectory, $\tilde{u}=\tilde{u}_0$ intersects the left quark trajectory at $x=x_0=-b\cosh\frac{\tilde{u}_0}{b}$. After the shock wave passes, for $\tilde{u}<\tilde{u}_0$, we assume the worldsheet is described by the embedding in $AdS_{d+1}$ as
  \be
  \label{bs}
  (x-c_x)^2-(t-c_t)^2=b_s^2-z^2 \, ,
  \ee
where $c_x$, $c_t$ and $b_s$ are constants. From the boundary point of view, we assume the left quark orginally with acceleration $1/b$ (following the trajectory $x=-\sqrt{t^2+b^2}$) emmits energy quanta at the proper time, $-\tilde{u}_0$ and changes the acceleration to $1/b_s$ with trajectory given by $x-c_x=-\sqrt{(t-c_t)^2+b_s^2}$. To match the worldsheet solutions for $\tilde{u}<\tilde{u}_0$ and $\tilde{u}>\tilde{u}_0$, we require that the quark trajectory and its first derivative are continuous across the boundary matching point $(x_0,t_0)$. These conditions gives,
 \be
  (x_0-c_x)^2-(t_0-c_t)^2-b_s^2=0,~~~\frac{(t_0-c_t)^2}{(t_0-c_t)^2+b_s^2}=\frac{t_0^2}{x_0^2} \, .
 \ee
Note that the derivative should be continuous, since the tangent line at the matching point is just the shock wave trajectory. These conditions uniquely determine $c_x$ and $c_t$ in region II,
 \be
 \label{cx}
 c_x=(b-b_s)\cosh\frac{\tilde{u}_0}{b},~~~c_t=(b-b_s)\sinh\frac{\tilde{u}_0}{b}\, .
 \ee
From (\ref{bs}), we can also parameterize this trajectory by the proper time $\tilde{u}_s$, so we have $x(\tilde{u}_s)-c_x=b_s\cosh(\frac{\tilde{u}_s}{b_s}+c_u)$, and $t(\tilde{u}_s)-c_t=b_s\sinh(\frac{\tilde{u}_s}{b_s}+c_u)$. Notice that the values of proper times, by definition, are the same at the matching point $\tilde{u}_s=\tilde{u}=\tilde{u}_0$. Thus the matching condition also determine the shifting in the origin of the proper time, $c_u=\tilde{u}_0(\frac{1}{b}-\frac{1}{b_s})$.

Even though the value of $\tilde{u}$ is continuous across the shock wave, the value of $\tilde{v}$ across the shock wave is shifted by some amount determined by the matching condition that the worldsheet needs to be continuous across $\tilde{u}=\tilde{u}_0$, that is the value of $z$ coordinate must match, which gives (see equation (\ref{z}))
 \be
 \label{match}
 b_s\tanh\frac{\tilde{u}_0-\tilde{v}_s}{2b_s}=b\tanh\frac{\tilde{u}_0-\tilde{v}}{2b} \, .
 \ee
Here $\tilde{v}_s$ is defined as in (\ref{uvti}) for region II, except using the shifted coordinates, $x-c_x$ and $t-c_t$. Note that the same matching condition is also obtained in \cite{Shenker_13} for the shock wave in a BTZ black hole. There, the role of $z$ is played by the radius of circle at the point $(\tilde{u},\tilde{v})$.

\begin{figure}[h]
\subfloat[Shock wave with positive $\gamma$. The wormhole remains non-traversable]{\label{fig1}\includegraphics[scale=0.8]{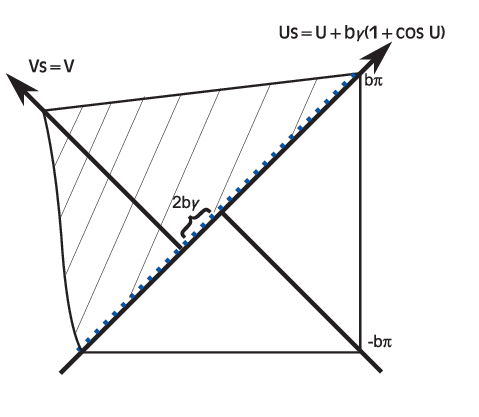}}
\subfloat[Shock wave with negative $\gamma$. The wormhole becomes traversable]{\label{fig2}\includegraphics[scale=0.8]{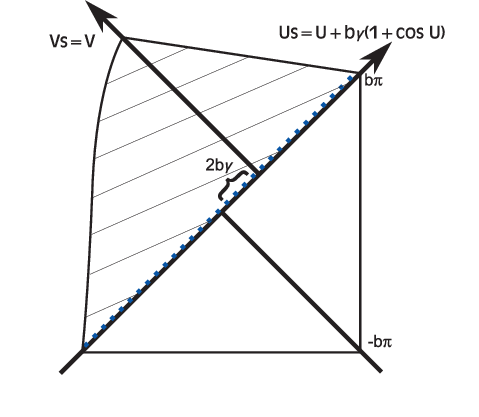}}
\caption{Shock wave backgrounds in double scaling limit with positive and negative $\gamma$ respectively. In both cases, the dash (blue) lines are the trajectories of the shock waves. The shaded regions are described by shifted coordinates $(U_s,V_s)$ (after the conformal transforms to match the values on the lines of the shock waves), and the unshaded regions are described by $(U,V)$-coordinate.}
\label{shockwaves}
\end{figure}

To simplify the expression for the following calculation of correlators, we consider the matching condition in the double scaling limit, where $b-b_s\rightarrow 0$ and $\tilde{u}_0\rightarrow\infty$ but keeping
\be
\frac1{2b}(b-b_s)e^{\frac{\tilde{u}_0}{b}}\equiv\gamma
\ee
fixed. Physically, we make the shock wave energy, $\delta E\propto b-b_s$ very small, and let the proper time ($-\tilde{u}_0$) for injecting energy very early.  Thus the blue-shift factor of the black hole, $e^{\tilde{u}_0/b}$ becomes very large and we keep the shock wave effect $\delta E e^{\tilde{u}_0}$ finite. In this limit, the matching condition (\ref{match}) gives
  \be
  \label{gamma}
  \tilde{v}_s=\tilde{v}+b\gamma e^{-\frac{\tilde{v}}{b}} \, .
  \ee
Thus the original horizon at $\tilde{v}=0$, is shifted by the amount $b\gamma$ after the shock wave passes.  It should be also noticed that the double scaling limit is not commuting with coordinate transformation. In different coordinate systems, the parameter $\gamma$ gets various kind of blue-shift (or red-shift) factors. For example, we can do the double scaling limit in the coordinate system (for region II)
\be
U'=-e^{\frac{\tilde{u}}{b}},~~V'=e^{\frac{-\tilde{v}}{b}} \, ,
\ee
then we have the discontinuity in $V'$ as
 \be
\label{gamma3}
 V'_s=V'+\gamma \, ,
 \ee
which agrees with the one in \cite{Shenker_13} and \cite{Murata_17}. Here it is more convenient to do the matching in $(U,V)$-coordinates defined above equation (\ref{UV}). The matching condition (\ref{match})(writing in this new coordinate system) in the double scaling limit gives
  \be
  \label{gamma2}
U_s=U+b\gamma(1+\cos\frac{U}{b}) \, .
  \ee
This relation is not the same as the relation (\ref{gamma}) or (\ref{gamma3}) in $(U,V)$-coordinate system. However if we take small $\gamma$ limit, then to leading order, this procedure gives the same relation as in (\ref{gamma}) or (\ref{gamma3}). Thus we will take (\ref{gamma3}) as a definition of $\gamma$, and equation (\ref{gamma2}) only the relation valid for small $\gamma$.
In the double scaling limit, we also have $c_x=c_t=-b\gamma$ and $c_u=0$ (see equation (\ref{cx}) and the following paragraph). The conformal diagrams for the shock wave backgrounds are drawn in figure \ref{shockwaves}. We see that for positive $\gamma$, the wormhole remains non-traversable, but some modes starting from $t=-\infty$, which originally ($\gamma=0$) can reach two quarks, may now go inside the horizon. And for negative $\gamma$ the shock wave makes the wormhole one-way traversable.

\section{real time correlators with a shock wave}
In this section, we use the method of holographic influence functional developed in \cite{Lee_15} to evaluate the field correlators that couple to quarks in this shock wave background discussed in previous section. We will first consider the linearized fluctuations on the shock wave background. The exact solutions of the linearized equation without shock wave have been found in \cite{Kawamoto_22}. Here we will do the matching of these solutions on shock wave surface in the double scaling limit to obtain the mode functions in the shock wave background. And then we will calculate the on-shell action and identify it as the generating functional for the fields that couple to the positions of the quarks. This allows us to derive the real-time field correlators. We will then discuss the behaviors of the correlators in various cases.

\subsection{Modes functions on the shock wave}
Here we consider the string worldsheet perturbations of the form $X^{\mu}=(t,z,x_b(t,z),q(t,z),0,\cdots, 0)$ where $q(t,z)$ is small as compared to $x_b(t,z)$. And the expansion of the Nambu-Goto action (\ref{action}) to the quadratic order gives
   \be
   \label{sq}
  S_q=-R^2T_0\int dt\, dz \frac{1}{z^2}\left(\frac{q'^2-\dot q^2-\dot x_b^2 q'^2-x'^2_b \dot q^2+2\dot x_b x'_b \dot q q'}{2\sqrt{1-\dot x_b^2+x_b'^2}}\right) \, .
  \ee
The equation of motion for $q(t,z)$ derived from this action can be solved exactly (see \cite{Kawamoto_22} for details). Here we write the general solutions found in \cite{Kawamoto_22} in the $(U,V)$-coordinate, which is defined below equation (\ref{uvprime}).
 \be
 \label{qi}
 q^i(U,V)=\int_{-\infty}^{\infty}\frac{d\omega}{2\pi}f^i(\omega)Q_{\omega}^i(U,V)+g^i(\omega)P_{\omega}^i(U,V) \, ,
 \ee
where $i=I,II,III,IV$, labels the four regions on the worldsheet. And we have
 \be
 \label{Qw}
 Q_{\omega}^I(U,V)=\frac{1-i\omega z(U,V)}{(1-i\omega z_m)e^{i\omega b\tanh^{-1}\frac{z_m}{b}}}e^{-i\omega b\ln\tan\frac{U}{2b}} \, ,
 \ee
and
 \be
 \label{Pw}
 P_{\omega}^I(U,V)=\frac{1-i\omega z(U,V)}{(1-i\omega z_m)e^{-i\omega b\tanh^{-1}\frac{z_m}{b}}}e^{i\omega b\ln\tan\frac{-V}{2b}} \, ,
 \ee
where $z(U,V)=b\frac{\cos\frac{U-V}{2b}}{\cos\frac{U+V}{2b}}$ is the value of $z$ in $(U,V)$-coordinate, and $z=z_m$ is the cut-off surface, where we impose the boundary conditions. Near the horizons $U=0$ and $V=0$ we have
 \bea
 &&Q_{\omega}^I(U\ll b,V)\rightarrow \frac{1-i\omega b}{(1-i\omega z_m)e^{-i\omega b\tanh^{-1}\frac{z_m}{b}}}\left(\frac{U}{2b}\right)^{-i\omega b}\equiv H(\omega)\left(\frac{U}{2b}\right)^{-i\omega b} \, , \label{Q1u}\\
 &&Q_{\omega}^I(U,V\ll b)\rightarrow  H(\omega)e^{-i\omega b\ln\tan\frac{U}{2b}}  \, ,\label{Q1v}
 \eea
and
  \bea
  &&P_{\omega}^I(U,V\ll b)\rightarrow H^*(\omega)\left(\frac{-V}{2b}\right)^{i\omega b}  \, ,\\
  &&P_{\omega}^I(U\ll b,V)\rightarrow H^*(\omega)e^{i\omega b\ln\tan\frac{-V}{2b}} \, .
  \eea
For the mode functions in region II, III and IV, we have the relations, $ Q_{\omega}^{II}(U,V)= Q_{\omega}^I(-U,-V)$, $ Q_{\omega}^{III}(U,V)= Q_{\omega}^{I}(U,-V)$ and $Q_{\omega}^{IV}(U,V)= Q_{\omega}^{I}(-U,V)$, and similarly for $P_{\omega}(U,V)$'s. In particular, in region III and IV, we have instead $z(U,V)=b\frac{\cos\frac{U+V}{2b}}{\cos\frac{U-V}{2b}}$.

We will now determine $q^i(U,V)$'s by their boundary conditions at the horizons ($U=V=0$) and the boundaries $z=z_m$, and thus fix $f^i(\omega)$'s and $g^i(\omega)$'s uniquely. At the boundaries we impose,
\be
\label{bczm}
q^I(z(U,V)=z_m)=q^R(t),~~ q^{II}(z(U,V)=z_m)=q^L(t) \, ,
\ee
where $q^R(t)$ and $q^L(t)$ are identified as the quark positions that source the boundary field operators, says $\hat F^R(t)$ and $\hat F^L(t)$.
We write
  \be
  \label{qaw}
  q^{a}(t)=\int_{-\infty}^{\infty}\frac{d\omega}{2\pi}\tilde{q}^a(\omega)e^{-i\omega \tau^a} \, ,
  \ee
where $a=L,R$ labels the right and left quarks. $\tau^L$ and $\tau^R $ are the proper times along the left and right quark trajectories respectively. They are related to the laboratory time $t$ by, $\tau^R=b\tanh^{-1}\frac{t}{\sqrt{t^2+b^2-z_m^2}}$ and $\tau^L=-b\tanh^{-1}\frac{t}{\sqrt{t^2+b^2-z_m^2}}$ (they are just the coordinate $\tilde{u}$ ($-\tilde{u}$) evaluated at the right (left) boundary as noticed earlier). Note that the normalization in (\ref{Qw}) and (\ref{Pw}) gives precisely $Q^I_{\omega}(z(U,V)=z_m)=P^I_{\omega}(z(U,V)=z_m)=e^{-i\omega \tau^R}$ and $Q^{II}_{\omega}(z(U,V)=z_m)=P^{II}_{\omega}(z(U,V)=z_m)=e^{-i\omega \tau^L}$. Thus the boundary conditions in (\ref{bczm}) give
\be
\label{bczm2}
f^{I}(\omega)+g^{I}(\omega)=\tilde{q}^R(\omega),~~~f^{II}(\omega)+g^{II}(\omega)=\tilde{q}^L(\omega) \, .
\ee

In the shock wave background after taking the double scaling limit as discussed in previous section, we have the shock wave following the line, $V=0$ and the coordinate $V$ is continuous across the shock wave, while $U$ is shifted by (\ref{gamma2}). As the consequence, the matching conditions for the fluctuations $q^{i}(U,V)$'s across the horizons are, firstly they should be analytic across $U=0$ (or $U_s=0$), and secondly they need to be matched across $V=0$ after the shifting of $U$ by (\ref{gamma2}). For example to find the matching conditions between region I and IV across $U=0$, we use the expression in (\ref{Q1u}) and that $Q_{\omega}^{IV}(U\ll b,V)\rightarrow H(\omega)\left(\frac{-U}{2b}\right)^{-i\omega b}$, which give
  \be
  \label{f14}
  f^I(\omega)=e^{-\pi b\omega}f^{IV}(\omega) \, .
  \ee
And near $U=0$, the similar expression for $P_{\omega}^{i}(U,V)$ gives the matching condition,
  \be
  \label{g14}
   g^I(\omega)=g^{IV}(\omega) \, .
  \ee
Note that the mode functions have branch cuts on complex $U$ and $V$ planes. We follow the prescription in \cite{Lee_19} to choose mode functions that are analytic in the upper (lower) half complex $U$ ($V$) plane. Thus to match $q^i(U,V)$'s across the shock wave, $V=0$ between region II and IV, we have
 \be
 \label{g24}
 g^{II}(\omega)=e^{-\pi b\omega}g^{IV}(\omega) \, .
 \ee
As for the matching of $f^{II}(\omega)$ and $f^{IV}(\omega)$, we use the expression (\ref{Q1v}) for region II and IV, and the shifting in (\ref{gamma2}), which give
 \be
 \label{fmatch}
\int_{-\infty}^{\infty}\frac{d\omega}{2\pi} f^{II}(\omega)H(\omega)e^{-i\omega b\ln\tan(\frac{-U}{2b}+\frac{\gamma}{2}(1+\cos\frac{-U}{b}))}=\int_{-\infty}^{\infty}\frac{d\omega}{2\pi}f^{IV}(\omega)H(\omega)e^{-i\omega b\ln\tan\frac{-U}{2b}}  \, .
 \ee
Use the small $\gamma$ expansion, $\ln\tan(\frac{-U}{2b}+\frac{\gamma}{2}(1+\cos\frac{-U}{b}))\simeq \ln\tan\frac{-U}{2b}+\gamma\cot\frac{-U}{2b}$, and note that in region II and IV, $\tilde{v}=b\ln\tan\frac{-U}{2b}$. This small $\gamma$ expansion is precisely the matching condition (\ref{gamma}) (see the discussion below equation (\ref{gamma2})). Thus to leading order in $\gamma$, we can write the exponents in the left hand side of (\ref{fmatch}) as
 \be
\ln\tan(\frac{-U}{2b}+\frac{\gamma}{2}(1+\cos\frac{-U}{b}))\simeq \tilde{v}+b\gamma e^{-\frac{\tilde{v}}{b}} \, .
 \ee
Multiply both sides of (\ref{fmatch}) by $e^{i\omega' \tilde{v}}$ and integrate over $\tilde{v}$, we obtain
 \be
 \label{f24integral}
f^{IV}(\omega')= f^{II}(\omega')+\frac1{H(\omega')}\int_{-\infty}^{\infty}\frac{d\omega}{2\pi}f^{II}(\omega)H(\omega)\Gamma(ib(\omega-\omega'))(-i\omega b \gamma)^{-ib(\omega-\omega')} \, .
 \ee
The integral is convergent for reasonable source functions. Note that the gamma function, $\Gamma(ib(\omega-\omega'))$ has poles on $\omega=\omega'+\frac{in}{b}$ for non-negative integers $n$'s. Thus to avoid the pole on real $\omega$ axis and make the integration well-defined, we use the standard $i\varepsilon$-prescription to make the pole at $n=0$ a little bit below the real axis, and after evaluating the integral by closing the integration contour on upper complex $\omega$ and picking up the residues of the poles, we take $\varepsilon\rightarrow0$. Using this procedure, we obtain
  \be
  f^{IV}(\omega')= f^{II}(\omega')+\frac1{H(\omega')}\sum_{n=1}^{\infty}f^{II}(\omega'+\frac{in}{b})H(\omega'+\frac{in}{b})\frac{(-i\gamma)^n}{n!}(b\omega'+ni)^n \, .
  \ee
This reminds us of the Matsubara frequency summation which reflects the underlying equilibrium system with temperature $T=\frac{1}{2b\pi}$. The inverse Fourier transform of the last term in the above equation can be written as (after doing the change of variable and the $n$-summation),
 \be
 \label{tb}
 \int_{-\infty}^{\infty}\frac{d\omega}{2\pi}f^{II}(\omega)(1-e^{-i\gamma b\omega e^{-\frac{t}{b}}})e^{-i\omega t} \, .
 \ee
In the latter discussion, we will be interested in two limits, $t\gg b$ and $t\ll b$ (take $t>0$). Thus, in the first limit, we obtain the matching condition
  \be
  f^{IV}(\omega)= f^{II}(\omega) \, ,
  \ee
which is the same as the one with no shock wave. This agrees with the finding in \cite{Shenker_13} that in the late times $t\gg b=\frac1{2\pi T}$, the effect of shock wave becomes negligible, since the laboratory frame in this limit is approximately the natural frame for infalling shock wave and there is no blue shift factor. In the opposite limit, $t\ll b$, we have
 \be
 \label{f24}
  f^{IV}(\omega)= f^{II}(\omega)(2-e^{-i\gamma b\omega}) \, ,
 \ee
if we can extrapolate the result to order-one $\gamma$. However we believe the result is only valid to leading order\footnote{There is another reason to believe that the result is only valid for small $|\gamma|$, which is that, for $|\gamma|>1$, the equation (\ref{gamma2}) becomes multi-valued.} in $\gamma$, then we have
 \be
 \label{f24s}
   f^{IV}(\omega)= f^{II}(\omega)(1+i\gamma b\omega) \, .
 \ee

\subsection{Holographic influence functional of the shock wave}
Now we are able to evaluate the on-shell action which we identify as the generating functional (or the influence functional) of boundary fields, and derive their correlators. From equations (\ref{f14}), (\ref{g14}), (\ref{g24}) and (\ref{f24s}), we see that the horizon boundary conditions give (using the identification $b=\frac{1}{2\pi T}$)
   \be
   f^{II}(\omega)=e^{\frac{\omega}{2T}}(1-\frac{i\gamma\omega}{2\pi T})f^{I}(\omega),~~~g^{II}(\omega)=e^{-\frac{\omega}{2T}}g^{I}(\omega) \, .
   \ee
Together with the boundary condition at $z=z_m$ in (\ref{bczm2}), we uniquely determine the solutions of $q^I(U,V)$ and $q^{II}(U,V)$. We then identify the on-shell action in (\ref{sq}), which contains only the boundary terms, as the generating function for the boundary fields $\hat F^L$ and $\hat F^R$, sourced by the quarks $q^L(t)$ and $q^R(t)$
 \bea
 \ln Z(\tilde{q}^L,\tilde{q}^R)&=&S_{q}(q^i(U,V))\nonumber\\
 &&=\int_{-\infty}^{\infty}\frac{d\omega}{2\pi}\tilde{q}^R(\omega)\left(\mbox{Re} G_R(\omega)-\frac{1-\frac{i\gamma\omega}{2\pi T}+e^{-\frac{\omega}{T}}}{1-\frac{i\gamma\omega}{2\pi T}-e^{-\frac{\omega}{T}}}i\mbox{Im} G_R(\omega)\right)\tilde{q}^R(-\omega)\nonumber\\
 &&+\tilde{q}^R(\omega)\left(\frac{2-\frac{i\gamma\omega}{\pi T}}{e^{\frac{\omega}{2T}}(1-\frac{i\gamma\omega}{2\pi T})-e^{-\frac{\omega}{2T}}}i\mbox{Im} G_R(\omega)\right)\tilde{q}^L(-\omega)\nonumber\\
 &&+\tilde{q}^L(\omega)\left(\frac{2}{e^{\frac{\omega}{2T}}(1-\frac{i\gamma\omega}{2\pi T})-e^{-\frac{\omega}{2T}}}i\mbox{Im} G_R(\omega)\right)\tilde{q}^R(-\omega)\nonumber\\
 &&+\tilde{q}^L(\omega)\left(-\mbox{Re} G_R(\omega)-\frac{1-\frac{i\gamma\omega}{2\pi T}+e^{-\frac{\omega}{T}}}{1-\frac{i\gamma\omega}{2\pi T}-e^{-\frac{\omega}{T}}}i\mbox{Im} G_R(\omega)\right)\tilde{q}^L(-\omega) \, ,\label{Zqq}
 \eea
where $G_R(\omega)\equiv \frac{R^2T_0}{2b^2}\frac{b^2-z_m^2}{z_m^2} Q_{\omega}^I(z=z_m)\partial_zQ_{\omega}^I(z=z_m)$, and is given by
\be
G_R(\omega)=\frac{R^2T_0}{2b^2}\left(\frac{b^2-z_m^2}{z_m}\frac{\omega^2}{1+\omega^2z_m^2}+i\frac{\omega+b^2\omega^3}{1+\omega^2z_m^2}\right) \, .
\ee
It is interpreted as the one-particle retarded Green function \cite{Kawamoto_22}. We then have the fields correlators
  \be
  G^{ab}(\omega)=\frac{\delta^2}{\delta \tilde{q}^a\delta \tilde{q}^b }S_{q}(q^i(U,V)) \, ,
  \ee
where $a,b=R,L$. It is easy to see that in the case $\gamma=0$, we have the (anti) time-order Green function for the $\hat F^R$ ($\hat F^L$) field in the thermal state with temperature $T$. Thus we regard $G^{ab}$ as the (anti) time-order Green functions in the back-reacted quantum state dual to the shock wave background. From the definition of the Fourier components in (\ref{qaw}), and note that $q^L$ is written in shifted coordinates (\ref{bs}) due to the shock wave, we then have
 \bea
 \label{grr}
 &&\langle\hat F^R(t)\hat F^R(t')\rangle=G^{RR}(t,t')=\frac{b}{\sqrt{b^2+t^2-z_m^2}}\frac{b}{\sqrt{b^2+t'^2-z_m^2}}\nonumber\\
 &&\times\int_{-\infty}^{\infty}\frac{d\omega}{2\pi}G^{RR}(\omega)e^{-i\omega ( \tau(t)- \tau(t'))} \, .
 \eea
And
 \bea
 \label{grl}
 &&\langle\hat F^R(t)\hat F^L(t')\rangle =G^{RL}(t,t')=\frac{b}{\sqrt{b^2+t^2-z_m^2}}\frac{b}{\sqrt{b^2+(t'+b\gamma)^2-z_m^2}}\nonumber\\
 &&\times\int_{-\infty}^{\infty}\frac{d\omega}{2\pi}G^{RL}(\omega)e^{-i\omega (\tau(t)+ \tau(t'+b\gamma))} \, ,
 \eea
where $\tau(t)=b\tanh^{-1}\frac{t}{\sqrt{b^2+t^2-z_m^2}}$. And $G^{LR}(t,t')$ and $G^{LL}(t,t')$ have the similar expressions.

\subsection{Time-dependence of the correlators}
The field correlators, $G^{ab}(t,0)$'s tell us how the effect of shock wave gets transmitted between quarks and finally thermalized and makes two quarks disentangled. We study the behavior of $G^{ab}(t,0)$ as $t\ll b=\frac1{2\pi T}$ (note that we also take $z_m$ as a UV-cutoff, and we consider $t\gg z_m$). Then the relevant $\omega$ in the integral is for $T\ll \omega\ll \frac1{z_m}$. In this limit, $G^{RR}(t,0)$ and similarly for $G^{LL}(t,0)$, reduces to the one for zero temperature time-order single particle correlator, since we have $G^{RR}(\omega)=\mbox{Re}G_R(\omega)-\mbox{sign}(\omega)\mbox{Im}G_R(\omega)$. In this limit, the time is too short or the wavelength of the fluctuations is too short to probe the interior of the black hole and the effect of the infalling shock wave. In this case, the time-dependent behavior of $G^{RR}(t,0)$ or $G^{LL}(t,0)$ has been studied in \cite{Lee_19}.

 \begin{figure}[h]
\centering
\includegraphics[scale=1]{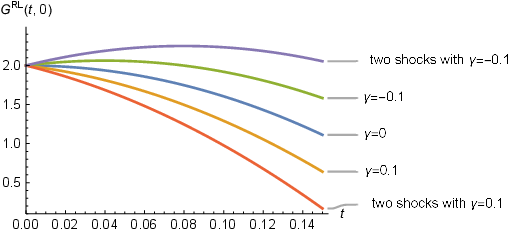}
\caption{The behavior of cross field correlator, $G^{RL}(t,0)$ for various shock wave configurations. Here $t$ is in unit of $\frac1T$. And $G^{RL}(t,0)$ is in unit of an overall constant as discussed in the text.}
\label{correlator}
\end{figure}

Thus to see the effects of the shock wave in the small $\gamma$ limit, we need to study the cross correlators $G^{RL}(t,0)$ or $G^{LR}(t,0)$\footnote{Two cross correlators are equal up to leading order in $\gamma$ as can be seen in (\ref{Zqq}).}. Here we focus on $G^{RL}(t,0)$. The integral in (\ref{grl}) is convergent and we can use the residues theorem to evaluate the integral. Note that the integrant has the poles at $\omega_n=-\frac{n\gamma}{2\pi b}+\frac{n i}{b}$(to leading order in $\gamma$) for integer $n$. Then in the limit $t\ll b=\frac1{2\pi T}$ and also taking $z_m\rightarrow 0$, we obtain, up to an overall time-independent constant, the leading time-dependent behavior of $G^{RL}(t,0)$ in small $\gamma$ and $Tt$ as
 \be
 \label{grlt}
 G^{RL}(t,0)\propto 2-10\pi \gamma T t-(2\pi Tt)^2 \, .
 \ee
We see that the cross correlation between left and right fields is decreasing eventually. This is expected since the entaglement between two quarks gradually lost, the fields coupled to them are also getting decoherent. In the case that $\gamma>0$, we can see that the effect of the shock wave is to increase the decay rate of the cross correlation. This is also expected, since the energy emitted by quarks disturbed the background further and that back-reacts to the quarks. An interesting observation is that when $\gamma$ is negative, equation (\ref{grlt}) tells us that the correlation actually increases for a while (as $t<\frac{5\gamma}{2\pi T}$), which is of the order of wormhole radius. We plot these behaviors of $G^{RL}(t,0)$ in figure \ref{correlator}(including the ones for two shock waves configuration discussed in next section). Usually we expect that, in field theory the energy released scrambles the ground state and decreases the correlation between two points, rather than builds up the correlation. But for this to happen, we need negative energy. This can be seen from the definition, $\gamma=\frac{b-b_s}{2b}e^{\frac{\tilde{u}_0}{b}}$ which is proportional to the shock wave energy (blue-shifted), and negative $\gamma$ implies negative energy for the shock wave, and it violates the null energy condition. So this kind of shock waves can't be produced by normal local perturbations on the worldsheet. As noted in \cite{Gao_19} (and in \cite{Boer_23} for the setup similar to current work), the negative energy shock waves on the worldsheet, for example, can be produced by introducing the non-local coupling between two quarks in the boundary theory,
 \be
 \label{doubletrace}
 \delta S=\int  h(t)q^L(-t)q^R(t) dt\, ,
 \ee
for some coupling function $h(t)$.

At the late times, when $t\gg\frac1T$, the effect from the shock wave is negligible (see equation (\ref{tb})), and the proper time $\tau(t)\simeq b\ln\frac{t}{b}$. We then have up to an overall constant
 \be
 G^{RL}(t,0)\simeq \frac1{t^2} \, .
 \ee
So the cross correlator decrease slower comparing to the one in early times, since the system is near the thermalized state.

\subsection{The relation to quantum chaos}
\label{chaos}
It is also interesting to evaluate $G^{RL}(0,0)$ and ask when the proper time $\tilde{u}_0$ is large enough to produce the order one effect on the cross correlator. This time scale, call it $\tilde{u}^*_0$ is defined as the scrambled time in \cite{Murata_17} and \cite{Shenker_13}. Here we can write
 \be
 \label{G00}
 G^{RL}(0,0)=\int_{-\infty}^{\infty}\frac{d\omega}{2\pi}G^{RL}(\omega)\mid_{\gamma=0}-\frac{i\gamma}{2\pi}\int_{-\infty}^{\infty}\frac{d\omega}{2\pi}\frac{\frac{\omega}{T}}{1-e^{-\frac{\omega}{T}}}G^{RL}(\omega)\mid_{\gamma=0} \, ,
 \ee
to leading order in $\gamma$, where $G^{RL}(\omega)\mid_{\gamma=0}$ is the correlator evaluate at $\gamma=0$. It is beyond our approximation here, when $\gamma$ is of order one. However we can still do the order of magnitude estimation of the time when the perturbation is of the order of the leading term. This happens when
 \be
 \gamma=\frac{b-b_s}{2 b}e^{\frac{\tilde{u}^*_0}{b}}\simeq \frac{\delta E}{E}e^{T\tilde{u}^*_0}\simeq 1 \, ,
 \ee
where $\delta E$ is the energy of the shock wave and $E$ is the energy of black hole with temperature $T=\frac1{2\pi b}$. We estimate that $E\simeq TS$ and $\delta E\simeq T$. Then we have the scramble time
  \be
  \tilde{u}^*_0=\frac1T\ln S \, ,
  \ee
which characterizes the thermalization rate of the black hole \cite{Sekino_08}.

The time-dependent correlators found here can be related to the analytic continuation of the out-of-time-order corelator (OTOC) and used to probe the quantum chaos\cite{Shenker_13,Maldacena_16}. This was noted in \cite{Roberts_14}, and in current setup, it was done in \cite{Murata_17,Boer_17,Banerjee_18,Banergee_19,Vegh_19,Kundu_21}. As suggested by the form of the influence functional in (\ref{Zqq}), without the shock wave($\gamma=0$), the left and right quantum fields that couple to two quarks can be described by the thermofield double (TFD) state with temperature $T$,
  \be
  |TFD\rangle=(Z_T)^{-1/2}\sum_n e^{-\frac{E_n}{2T}}|n\rangle_L\otimes |n\rangle_R \, ,
  \ee
where we assume the quantum fields have discrete energy spectrum and $Z_T$ is the normalization factor. The emission of the shock wave at the time $t_0$ from the left quark can be described by some operator $\hat W_L(t_0)=\hat W(t_0)\otimes\hat 1$ acting on $|TFD\rangle$. We can then identify, for example,
 \be
 G^{RL}(0,0)=\langle TFD|\hat W_L^{\dagger}(t_0)\hat F^R(0)\hat F^L(0)\hat W_L(t_0)|TFD\rangle \, .
 \ee
If we further define $\hat F^L(t)=\hat F(t)\otimes \hat 1$ and $\hat F^R(t)=\hat 1\otimes \hat F^T(t)$, where $\hat F^T$ is the transpose of $\hat F$, then it is not difficult to see that $G^{RL}(0,0)$ is equal to an analytic continuation of the four-point thermal OTOC,
  \be
  \label{otoc}
  \langle \hat W^{\dagger}(t_0)\hat F(0)\hat W(t_0)\hat F(\frac{i}{2T})\rangle_T \, ,
  \ee
where $\langle\hat O\rangle_T=Tr(\hat O\hat \rho_T)$ with $\hat \rho_T=Z_T^{-1}\sum_n e^{-\frac{E_n}{T}}|n\rangle\langle n| $. Thus from the expression for $G^{RL}(0,0)$ in (\ref{G00}) we can also read off the OTOC, which scales with $\tilde{u}_0$ as $1-constant\times e^{\frac{\tilde{u}_0}{b}}$. Thus it exhibits a chaotic behavior with Lyapunov exponent $\lambda_L=2\pi T$, which saturates the bound in \cite{Maldacena_16}.  This agrees with the result in \cite{Murata_17}. Noted that $-\tilde{u}_0$ is the proper time of the quark, and it is related to the laboratory time, $t_0=b\sinh\frac{-\tilde{u}_0}{b}$. Thus the exponential growth of the four-point OTOC is only observed in the frame when quarks see the thermal bath, and there is no such growth in the laboratory frame. Also it should be noted that in current work, the expression in (\ref{G00}) is only valid in small $\gamma$, thus the comparison to the results in \cite{Murata_17,Boer_17,Banerjee_18,Banergee_19,Vegh_19,Kundu_21} is only qualitative. It will be interesting to extend the result to finite $\gamma$ to see the general dependence of the OTOCs on $\tilde{u}_0$. We will discuss more about this point in the final section.

\section{two shock waves}
\label{shocks}
\begin{figure}[h]
\centering
\includegraphics[scale=1]{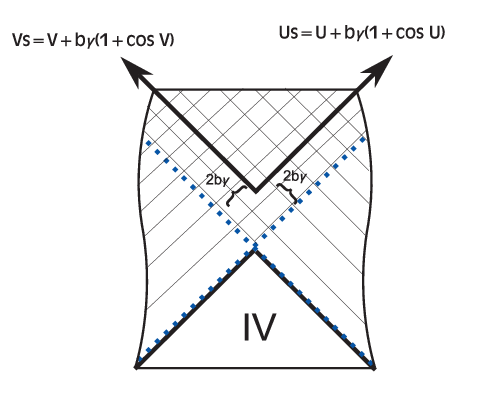}
\caption{The worldsheet conformal diagram for two shocks from left and right boundaries in the double scaling limit. The shock wave trajectories are described by dash (blue) lines. Here we draw only the case when $\gamma$ is negative, so a two-way traversable wormhole is formed. The unshaded region (region IV) is described by $(U,V)$-coordinate. The shaded regions are described by shifting one of $U$ and $V$ to $U_s$ or $V_s$. The double shaded region is described by $(U_s,V_s)$-coordinate.}
\label{twoshocks}
\end{figure}

We now consider the situation when two quarks emit energy quanta simultaneously from the far past toward each other. In the double scaling limit and with $\gamma$ negative, we plot the worldsheet conformal diagram in figure \ref{twoshocks}.

In the situation with two shock waves on the worldsheet, by the similar consideration in previous section, we obtain the following matching conditions in the early limit ($t\ll \frac1T$)
     \be
   f^{II}(\omega)=e^{\frac{\omega}{2T}}(1-\frac{i\gamma\omega}{2\pi T})f^{I}(\omega),~~~g^{II}(\omega)=e^{-\frac{\omega}{2T}}(1+\frac{i\gamma\omega}{2\pi T})g^{I}(\omega) \, .
   \ee
Then we have the correlators in Fourier space as
  \bea
  &&\tilde{G}^{RR}(\omega)=\mbox{Re}G_R(\omega)-\frac{1-\frac{i\gamma\omega}{2\pi T}+e^{-\frac{\omega}{T}}(1+\frac{i\gamma\omega}{2\pi T})}{1-\frac{i\gamma\omega}{2\pi T}-e^{-\frac{\omega}{T}}(1+\frac{i\gamma\omega}{2\pi T})}i\mbox{Im} G_R(\omega) \, ,\\
  &&\tilde{G}^{RL}(\omega)=\frac{2(1-\frac{i\gamma\omega}{2\pi T})(1+\frac{i\gamma\omega}{2\pi T})}{e^{\frac{\omega}{2T}}(1-\frac{i\gamma\omega}{2\pi T})-e^{-\frac{\omega}{2T}}(1+\frac{i\gamma\omega}{2\pi T})}i\mbox{Im} G_R(\omega) \, .
  \eea
And the expressions for $LL$ and $LR$ components are similar. Comparing with (\ref{Zqq}), we find that to leading order in $\gamma$, $\tilde{G}^{ab}(\omega)$ is just equal to $G^{ab}(\omega)$ by replacing $\gamma$ by $2\gamma$. In particular, in the case with two shock waves, we have the time-dependent cross correlator, at the early times $Tt\ll 1$, as
 \be
 \tilde{G}^{RL}(t,0)\propto 2-20\pi \gamma T t-(2\pi Tt)^2 \, .
 \ee
Thus when $\gamma>0$ the amount of decreasing in the cross correlation is double comparing to the case with one shock wave. When $\gamma<0$ we now have a two-way traversable wormhole, and the time interval for the cross correlation to build up (now for $t<-\frac{5\gamma}{\pi T}$) is also double comparing to the case with one-way traversable wormhole.

\section{discussion}
So far we haven't discussed the relation between shock wave backreation and its effect on entanglement entropy of the EPR pair. As we studied in \cite{Lee_19}, the dynamics of quarks is largely determined by the field-field correlators $\langle \hat F^a(t) \hat F^b(t')\rangle$. Thus, there is good reason to believe that the cross correlator we found in (\ref{grlt}) captures qualitatively the interaction between two entangled quarks and may tell us how the entanglement get changed due to the shock waves. However the detailed time-dependence can only be obtained by solving the coupled Lagenvin equations between two quarks, similar to the formula obtained in \cite{Syu_21}.

We found that in the limit $t\ll \frac1T$, when $\gamma$ is positive, the cross field correlators decrease more quickly than the case $\gamma=0$. And in case when $\gamma$ is negative, the wormhole is traversable, and we have the cross field correlation builds up for a while, and then after $t\simeq\gamma\frac1T$, which is of the order of the wormhole throat size, it starts to decrease. These behaviors of cross correlators of fields have a natural explanation in the ER=EPR conjecture, if we assume they also capture qualitatively the quarks dynamics.  As $\gamma>0$, even though from the worldsheet point of view, two quarks remain causally disconnected, more information (on the initial state) is leaking to the regions behind the horizons and that should make the correlation between quarks decrease faster. This is also consistent with the interpretation given in \cite{Pedraza_13}, which identified the regions behind the horizons to be the one in gravity theory to encode the information of gluon fields. So from the boundary point of view, the quanta sending between the EPR quark pair, dissipate the information to gluons and make the quark pair less entangled. And as $\gamma<0$, the wormhole is rendered traversable by shock waves. But from the boundary point of view, two quarks are still casually disconnected. So it looks like the superluminal communication between quarks is possible, and that communication builds up the correlation between quarks. This may also relate to the fact that the negative energy shock waves ($\gamma<0$) is ultimately attributed to the non-local interaction in the theory(for example the double trace deformation as described in (\ref{doubletrace})). It will be very interesting to see if the time evolution of entanglement entropy between two entangled quarks behaves as expected. We plan to study this in future works.

Another interesting direction for the future works is the study of quantum chaos by OTOCs as mentioned in section \ref{chaos}. Comparing to other holographic calculations of OTOCs which are based on the Euclidean holography, we work on Lorentzian time directly. Since there is no need to do the analytic continuation to real times, the numerical methods is more easily implemented in the holographic influence functional. Also as mentioned in the introduction already, the holographic influence functional method makes it easier to study the quark dynamics, and we may calculate the correlators in different vacuum states by changing the boundary conditions of the bulk mode functions at the horizons\cite{Lee_15}. However, the difficulty to find the general analytic form of OTOCs in current work is to go beyond the small $\gamma$ approximation of the correlators. Also, as can be seen in (\ref{tb}), in general (not just in the early time and late time limits considered here), the bulk mode functions with different frequency coupled together and we have more complicated influence functional. If this difficulty can be overcome, we can also study the scattering in the wormhole interior as in \cite{Haehl_2104,Haehl_2105}, which is related to the two shock waves backgrounds discussed here in section \ref{shocks}. From the boundary point of view, the correlators, $\tilde{G}^{RL}(0,0)$, we found here can be related to an analytic continuation of the six-point OTOC of the type, $\langle \hat W^{\dagger}(t_1)\hat F(t_2)\hat W(t_3)\hat W^{\dagger}(t_4)\hat F(t_5)\hat W(t_6)\rangle_T$, which is the one studied in \cite{Haehl_2104,Haehl_2105}.

\acknowledgments

I am grateful to D.-C. Dai, D.-S. Lee, Y. Sekino, and K. Shoichi for useful discussions and comments on this manuscript. The work was supported in part by NSTC grant 112-2112-M-259-016.

\end{document}